\documentclass[onecolumn,english,groupedaddress, superscriptaddress,aps,pre,10pt,reprint]{revtex4-2}

\usepackage[T1]{fontenc}
\usepackage[utf8]{inputenc}

\usepackage[colorlinks=true,linkcolor=blue,urlcolor=blue,citecolor=blue,breaklinks=true]{hyperref}
\usepackage{xurl}

\usepackage{amsmath}

\usepackage{graphicx}

\usepackage{siunitx}

\begin{document}
\title{Initial analysis of the impact of the Ukrainian power grid synchronization with Continental Europe}

\author{Philipp~C.~Böttcher}
\thanks{Shared authorship}
\affiliation{Forschungszentrum Jülich, Institute for Energy and Climate Research - Systems Analysis and Technology Evaluation (IEK-STE), 52428 Jülich,
Germany}

\author{Leonardo~Rydin~Gorj\~ao}
\thanks{Shared authorship}
\affiliation{Department of Computer Science, OsloMet -- Oslo Metropolitan University, N-0130 Oslo, Norway}

\author{Christian~Beck}
\affiliation{Queen Mary University of London, School of Mathematical Sciences, Mile End Road, London E1 4NS, UK}
\affiliation{The Alan Turing Institute, 96 Euston Road, London NW1 2DB, UK}

\author{Richard~Jumar}
\affiliation{Institute for Automation and Applied Informatics, Karlsruhe Institute of Technology, 76344 Eggenstein-Leopoldshafen, Germany}

\author{Heiko~Maass}
\affiliation{Institute for Automation and Applied Informatics, Karlsruhe Institute of Technology, 76344 Eggenstein-Leopoldshafen, Germany}

\author{Veit~Hagenmeyer}
\affiliation{Institute for Automation and Applied Informatics, Karlsruhe Institute of Technology, 76344 Eggenstein-Leopoldshafen, Germany}

\author{Dirk~Witthaut}
\thanks{Shared authorship}
\affiliation{Forschungszentrum Jülich, Institute for Energy and Climate Research - Systems Analysis and Technology Evaluation (IEK-STE), 52428 Jülich,
Germany}
\affiliation{Institute for Theoretical Physics, University of Cologne, 50937 Köln,
Germany}

\author{Benjamin Schäfer}
\thanks{Shared authorship}
\affiliation{Institute for Automation and Applied Informatics, Karlsruhe Institute of Technology, 76344 Eggenstein-Leopoldshafen, Germany}

\begin{abstract}
When Russia invaded Ukraine on the  24\textsuperscript{th} of February 2022, this led to many acts of solidarity with Ukraine, including support for its electricity system. 
Just 20 days after the invasion started, the Ukrainian and Moldovan power grids were synchronized to the Continental European power grid to provide stability to these grids.
Here, we present an initial analysis of how this synchronization affected the statistics of the power grid frequency and cross-border flows of electric power within Continental Europe.
We observe faster inter-area oscillations, an increase in fluctuations and changes in the cross-border flows in and out of Ukraine and surrounding countries as an effect of the synchronization with Continental Europe.
Overall these changes are small such that the now connected system can be considered as stable as before the synchronization.
\end{abstract}
    
\maketitle

\section*{Introduction}
On the 24\textsuperscript{th} of February 2022, Russian armed forces invaded Ukraine in violation of the Law of Nations.
The ongoing conflict has far-reaching consequences for the Ukrainian energy system.
The case of natural gas supply is probably best known and has been heavily discussed in the literature, see, e.g.~\cite{bradshaw2009geopolitics,kutcherov2020russian} and references therein.
For decades, Ukraine has been dependent on the supply of natural gas~\cite{wolczuk2016managing} but Russia repeatedly curtailed supplies, for instance during the Russo-Ukrainian gas dispute in 2009~\cite{stern2009russo}.

The Russian invasion also had implications for the European and Ukrainian electric power systems. Historically, most of the Ukrainian power grid was in synchronous operation with the interconnected network of the Community of Independent States (CIS) \cite{yandulskyi2018analysis}, featuring Russia as the largest country.
Due to the ongoing conflicts, perspectives for an integration with the Continental European power systems were evaluated already in 2017~\cite{entsoe2017agreement} and 2021~\cite{feldhaus2021connecting}.
At the same time, tensions were rising between Russia and Moldova, whose grid was also synchronous to the CIS interconnected networks.
As a consequence, transmission system operators were striving to synchronize with the Continental Europe power system for years.
On the 16\textsuperscript{th} of March, just 20 days after the invasion, an emergency synchronization was carried out~\cite{entsoe2022sync}.
While preparation has been going on for years, the final steps were realized very quickly and it remains unclear how such a  synchronization impacts the Continental European power grid.

Within this article we investigate the impact of this emergency synchronization on the operation of the Continental European power grid from a statistical perspective.
After providing some further background information, we first analyze recordings of the power-grid voltage frequency (in the following simply referred to as frequency) and secondly evaluate cross-border flows of electric power, aiming to quantify changes in the statistical properties of power grid operation.
Overall, we find small but notable changes in the operation due to the synchronization.

\section*{Background}

The stable operation of a power grid requires that all generators run in synchrony at a common frequency~\cite{witthaut2022collective}.
Transient violations of perfect synchronization can occur after local disturbances~\cite{Klein1991}, but perturbations are rapidly damped out~\cite{rydingorjao2020open}. 
In contrast, a full loss of synchrony can have fatal consequences.
For instance, the Continental European power grid split into three mutually asynchronous fragments during the 2006 European power outage after a cascade of transmission line failures~\cite{UCTE07}. 
Similar, although not as severe, system split events took place in 2021~\cite{entsoe2021a, entsoe2021b}.

Several synchronous areas exist in Europe, with the Continental European (CE) area being the largest.
Prior to 2022, it included a generation capacity of more than $600 \, \si{GW}$ and served more than 400 million customers.
Power transmission between different synchronous areas is possible only via converters~\cite{kolar2011review} or high-voltage directed current (HVDC) lines~\cite{pierri2017challenges}.
Regulations for power system operation and control are standardized within the European Network of Transmission System Operators for Electricity (ENTSO-E). 

The Moldovan power grid and most of the Ukrainian power grid were part of the Integrated/Unified Power System (IPS/UPS), a large synchronous area, spanning most of the Community of Independent States (CIS)~\cite{yandulskyi2018analysis}.
In this grid, load-frequency control is organized by the Russian grid operator~\cite{feldhaus2021connecting}.
An exception was the Burshtyn Island in South-Western Ukraine, which has been isolated from the rest of the Ukrainian grid and had an interconnection with the ENTSO-E~\cite{yandulskyi2018analysis}.
Notably, the Baltic states are still in synchronous operation with the IPS/UPS but are working to leave this area and synchronize with the CE grid instead~\cite{bompard2017electricity,entsoe2018european,putkonen2022modeling}.

First plans to connect Ukraine to the Continental European power grid have been formulated in a memorandum of understanding between the European Union and Ukraine in 2005 and reconfirmed in 2016~\cite{eu2005memorandum,feldhaus2021connecting}.
On the 28\textsuperscript{th} June, 2017, transmission system operators from Continental Europe, Ukraine and Moldova signed the ``Agreements on the Conditions of the Future Interconnection of the Power System of Ukraine and Moldova with the Power System of Continental Europe''~\cite{entsoe2017agreement}. 
This agreement has specified several technical requirements and security measures to be implemented prior to a synchronization with the CE grid.
Three days after the Russian invasion, on the 27\textsuperscript{th} February, 2022, Continental Europe Transmission System Operators (TSOs) received an urgent request from Ukrenergo, the Ukrainian TSO, for an emergency synchronization of the Ukrainian power system, including the Burshtyn island, followed by a similar request by Moldelectrica, the Moldovan TSO, one day later~\cite{entsoe2022request}. 
This emergency synchronization was carried out on the 16\textsuperscript{th} of March, 2022, 20 days after Russia started the invasion of Ukraine~\cite{entsoe2022sync}, fundamentally ``doing a year's work in two weeks'' as phrased by the European Commission~\cite{eu2022statement}.

\section*{Statistics of power system frequency data}

The grid frequency indicates the balance of power in the grid: It decreases in the case of scarcity and increases in the case of an oversupply. 
Hence, frequency is the prime observable in power system control~\cite{witthaut2022collective}.
Primary control, also referred to as frequency containment reserve (FCR) in Europe, is provided by specific power plants, hydropower stations or batteries.
It is activated within seconds and adapts power generation proportional to the frequency apart from a small dead band~\cite{handbook2009policy, handbook2004appendix}.
Secondary control, also referred to as frequency restoration reserve in Europe, restores the grid frequency to its set value and reduces unscheduled power flows between different areas in the grid~\cite{handbook2009policy, handbook2004appendix, Fleer2018, Ullah2021}.
Within this section, we investigate how the frequency statistics but also derived quantities, such as the estimated control amplitude changed due to the synchronization of Ukraine and Moldova.
The data we use here was recorded using phasor measurement units (PMUs) kindly provided by Gridradar, having a temporal resolution of $100$\,ms~\cite{gridradar}.

\subsection*{Elementary statistic characterization}

\begin{figure}[t]
    \begin{center}
    \includegraphics[width=0.9\columnwidth]{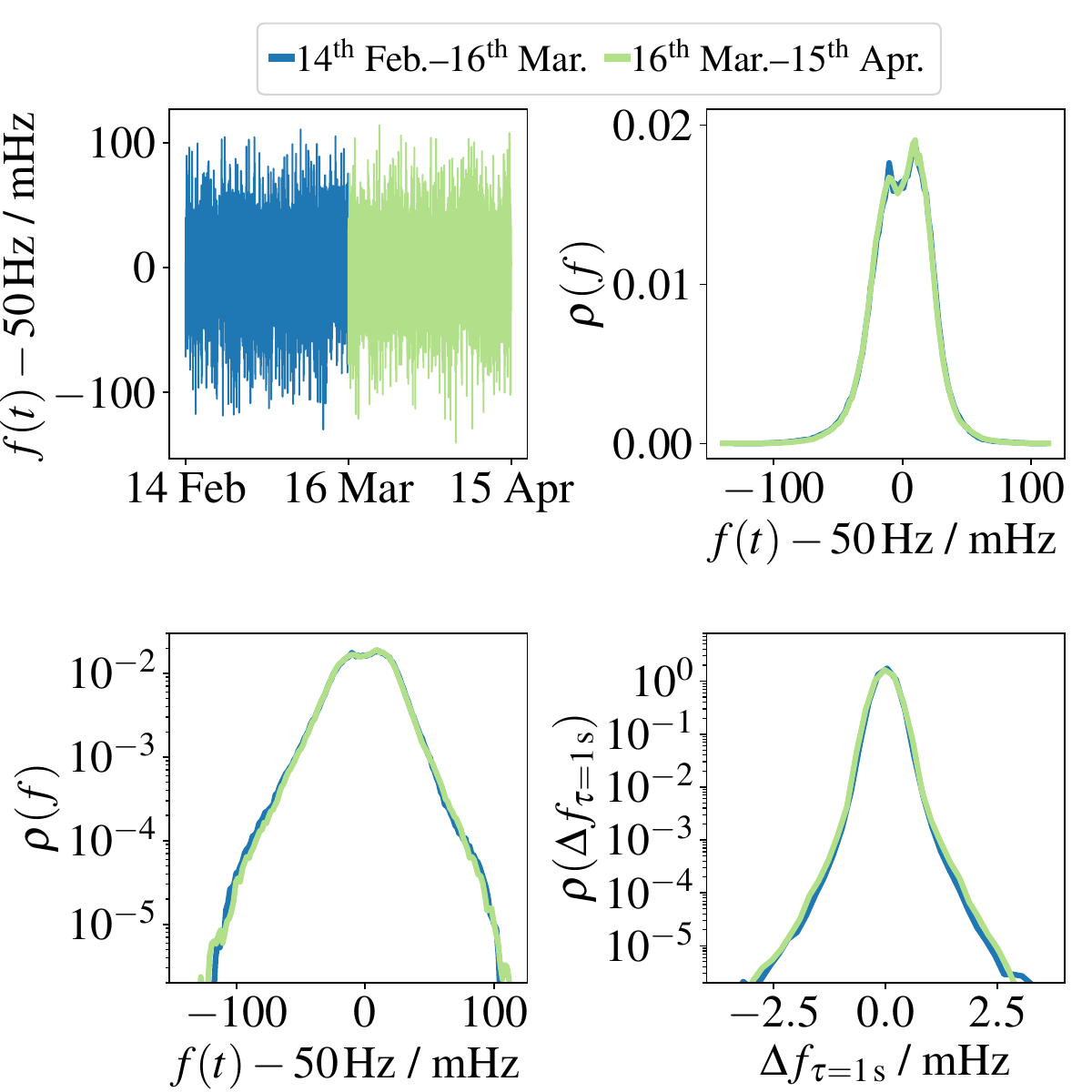}
    \end{center}
    \caption{Trajectory and probability density of the power-grid frequency recordings four weeks prior and four weeks past the 16\textsuperscript{th} of March. 
    Top Left: Trajectory of the frequency recordings as measured in Bremen, Germany.
    Top Right: Probability density function (PDF) of the trajectories prior and post connection.
    The distinct double peak is due to the deadband in the FCR control law in the interval $[-10,10]$ mHz around the set value.
    Bottom Left: PDF of the trajectories on a logarithmic scale.
    Bottom Right: PDF of the frequency increments $\Delta f_{\tau}(t) = f(t+\tau) - f(t)$ for $\tau = 1 \si{s}$ on a logarithmic scale.}
    \label{fig:1}
\end{figure}

\begin{figure*}[t]
    \begin{center}
       \includegraphics[width=0.9\textwidth]{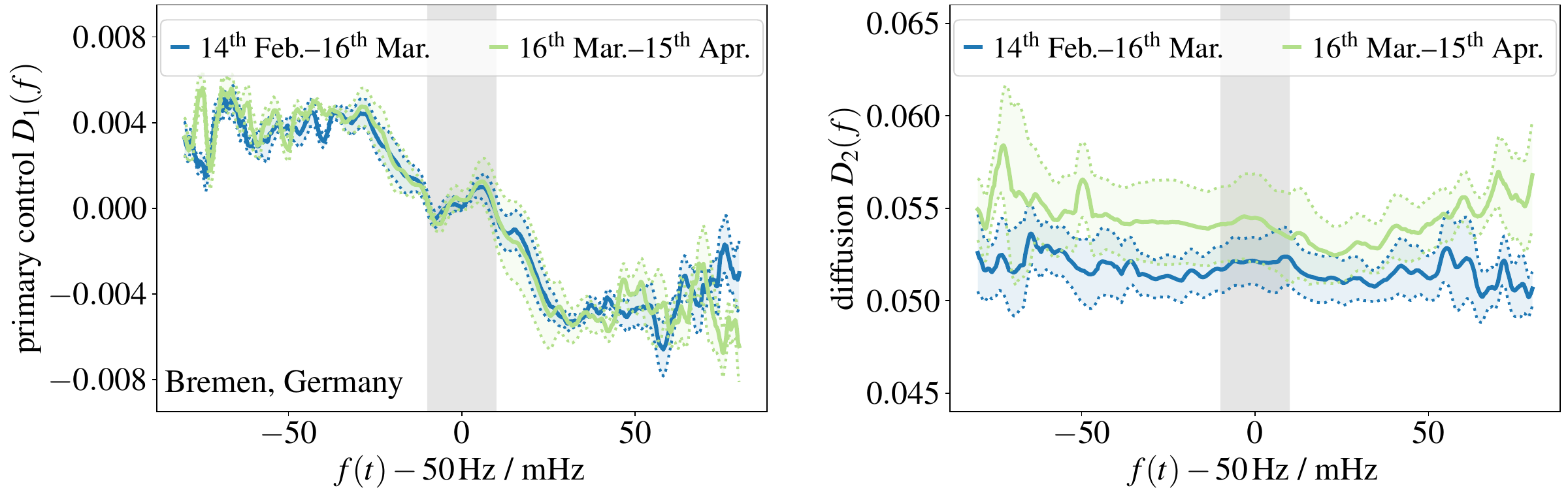} 
    \end{center}
    \caption{Drift and diffusion of power-grid frequency recordings four weeks prior and four weeks past the 16\textsuperscript{th} of March. 
    (left panel) The drift $D_1(f)$ of the recordings, which is an indicator for the amount of control in the system, i.e., the linear response to deviations from the nominal frequency of $50$~Hz. 
    We observe essentially no change in the control before and after the synchronization. 
    (right panel) The diffusion $D_2(f)$ of the recordings, which shows a large amount of `noise' after the 16\textsuperscript{th} of March. 
    This indicates that the integration of the Moldovan-Ukrainian grid into the CE grid increased the overall noise in the grid.
    The dotted lines and shaded areas indicated a $\pm 1$ standard deviation from bootstrapping the analysis into contiguous segments of one week from the starting date, the 14\textsuperscript{th} of February.
    All figures were produced with \texttt{Matplotlib}, \texttt{NumPy}, \texttt{SciPy}, and \texttt{kramersmoyal} ~\cite{Matplotlib, NumPy, SciPy, rydingorjao2019kramersmoyal}.}
    \label{fig:2}
\end{figure*}

As a first, straight-forward analysis, we compare the statistics of the grid frequency $f(t)$ in the Continental European synchronous area four weeks prior and post the synchronization of Ukraine and Moldova on the 16\textsuperscript{th} of March, see Fig.~\ref{fig:1}. 
The probability density of $f(t)$ shows a characteristic double peak. 
This is due to a deadband in the FCR control law, where no control power is applied if the frequency remains within an interval of $[-10,+10] \, \si{mHz}$ around the set frequency of $50 \, \si{Hz}$.
The PDF decays approximately exponentially for deviations far away from the set frequency (in the tails of the distribution).

The comparison of the time periods before and after the 16\textsuperscript{th} of March shows only very small differences. 
Overall, the are no signs for a general decrease of frequency stability. 
Further data are needed to make definite statements regarding the probability of extreme events at the tails of the distribution.

In addition, we analyze the increments of the frequency 
\begin{equation}
    \Delta f_{\tau}(t) = f(t+\tau) - f(t),
\end{equation}
which provide information about short-term power imbalances in the grid \cite{anvari2020stochastic,rydin2021spatio}.
In fact, a positive value corresponds to a temporary oversupply of power that accelerates rotating synchronous machines, while a negative value corresponds to a scarcity of power.
Even when evaluating only four weeks, we note that the PDF of the increments for $\tau = 1 \, \si{s}$ is not fully Gaussian but large increments are observed more frequently \cite{rydingorjao2020open}. 
Differences between the two periods are again very small. We might speculate that the values at the tails, i.e. particularly rare but large deviations, occur slightly more often after the synchronization.

\subsection*{Drift and diffusion}

For a more specific analysis, we model the frequency trajectory $f(t)$ as a stochastic process.
On coarse-scales the evolution of the grid frequency is described by the equation~\cite{schafer2018non,rydingorjao2020data}
\begin{align*}
    M \frac{\mathrm{d}f}{\mathrm{d}t} &=  \Delta P_{\rm tot}(t), 
\end{align*}
where $M$ is the aggregated inertia and $\Delta P_{\rm tot}(t)$ the total short term power imbalance.
In the subsequent analysis, we decompose the imbalance into three contributions: 
A slowly varying systematic power imbalance $\Delta P(t)$, the primary control of FCR that depends of the frequency, and zero-mean high-frequency fluctuations. 
Dividing by the inertia, we thus have the stochastic differential equation
\begin{align*}
    \frac{\mathrm{d}f}{\mathrm{d}t} = \Delta P(t) + D_1(f)  +  D_2(f) \xi(t),
\end{align*}
where $\xi(t)$ is uncorrelated stochastic noise.
Systematic power imbalances $\Delta P(t)$ arise for instance due to the rapid ramping of generators at the beginning of each hour causing deterministic frequency deviations~\cite{Weissbach2009,kruse2021exploring}. 
Furthermore, we absorb the effect of the frequency restoration reserve in this term.

The functions $D_1(f)$ and $D_2(f)$ can be inferred from data via the Kramers--Moyal expansion~\cite{schafer2018non,rydingorjao2020data}.
Results are shown in Fig.~\ref{fig:2}. 
Except for the tails, where data are scarce, the functions $D_1(f)$ approximately matches the expected proportional control law with a deadband:
\begin{equation}
    D_1(f) = \left\{ \begin{array}{l c l}
    - \frac{c_1}{M} (f - f_-) & &   f < f_- , \\
    0 & \mbox{ for } & f_- \le f \le f_+, \\
    - \frac{c_1}{M} (f - f_+) & & f_+ < f,
    \end{array} \right.
\end{equation}
where $[f_-,f_+] = [49.99,50.01] \, \si{Hz}$ is the deadband.
We recall that we have divided the equations of motion by the inertia $M$, such that the proportionality constant is the aggregated droop constant $c_1$ divided by $M$.

The data suggests that the ratio $c_1/M$ remained approximately the same as before the synchronization on the 16\textsuperscript{th} of March. 
We can assume that the effective inertia $M$ increased as the power plants in Ukraine and Moldova synchronized with the CE grid, especially given that these regions rely on nuclear and fossil fuel power plants~\cite{BPreview2021}.
Hence, the aggregated droop constant $c_1$ has increased proportionally.

The diffusion function $D_2(f)$ is approximately constant with respect to $f$, showing that the power fluctuations are mostly independent of the actual frequency $f(t)$.
We find that the strength of the fluctuations slightly \emph{increased} after the synchronization with Ukraine on 16\textsuperscript{th} of March.
This finding is consistent with the idea that the frequency increments shown in Fig.~\ref{fig:1} show large deviations more often.

These observations suggest the following interpretation.
The overall inertia of the CE grid increased due to the considerable amount of conventional generation in Ukraine and Moldova~\cite{BPreview2021}.
This would theoretically decrease the amplitude of the diffusion $D_2$.
Contrary to this expectation, we observe a slight increase in the amplitude of diffusion, which points to added fluctuations in the system.
Yet it does not seem that the overall frequency quality has changed substantially.

\subsection*{Periodic modes}

\begin{figure*}[tb]
    \centering
    \includegraphics[width=.62\textwidth]{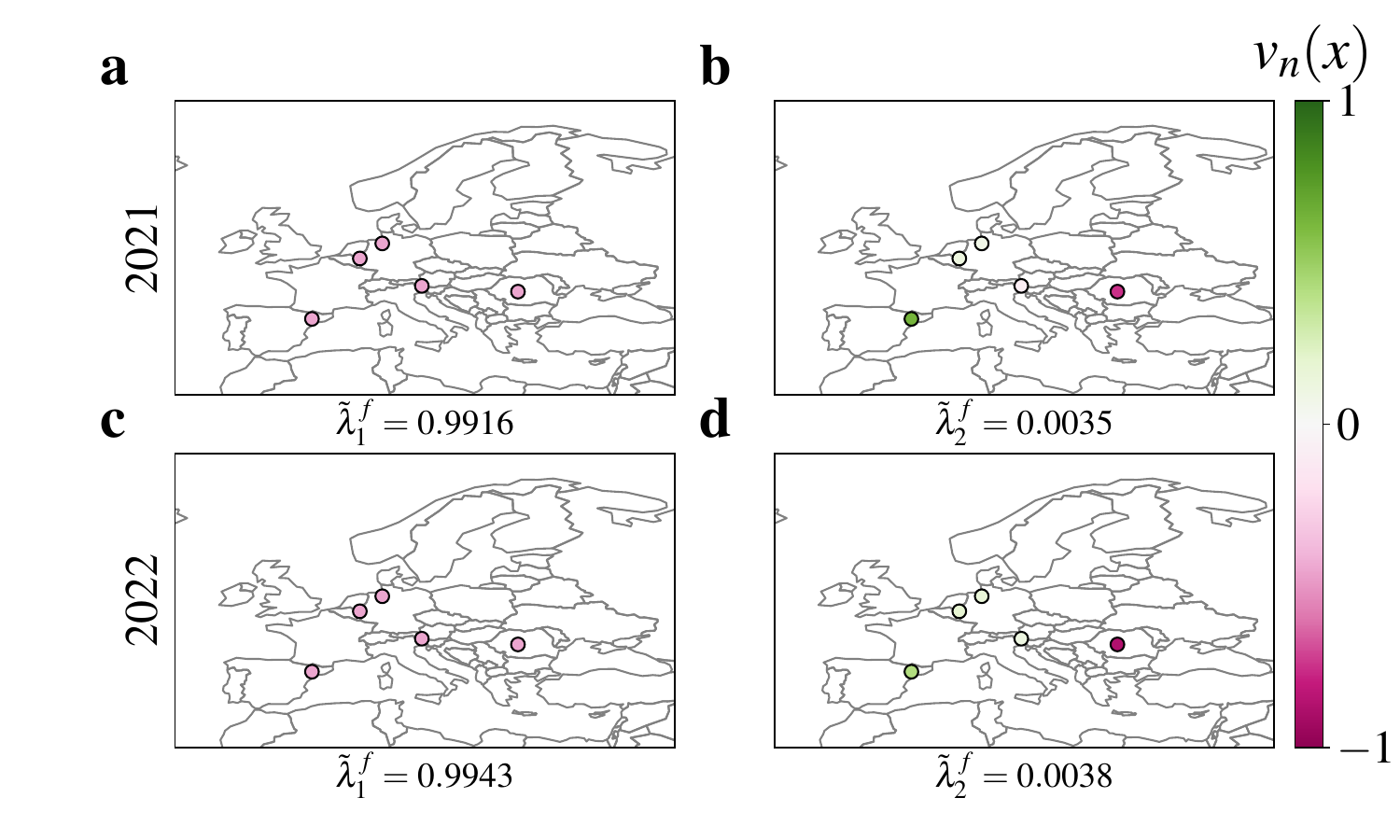}
    \includegraphics[width=.29\textwidth]{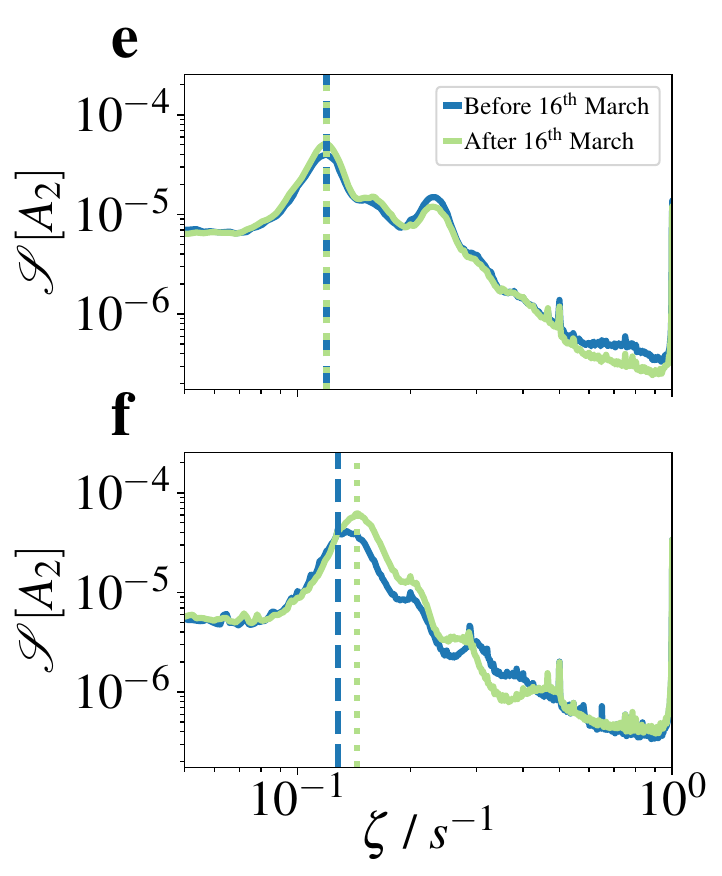}
    \caption{
    Inter-area modes in the Continental European grid analyzed by Principal Component Analysis (PCA). 
    Panels \textbf{a-d} show the two dominant principal components (PCs) $v_n(x)$ obtained from synchronized frequency measurements at five locations in 2021 (a,b) and 2022 (c,d).
    The first PC (a,c) corresponds to the bulk grid dynamics, while the second PC (b,d) corresponds to the East-West inter-area mode. 
    Panels \textbf{e} and \textbf{f} show the power spectral density (PSD) $\mathcal{S}$ of the amplitude of the second PC $A_2(t)$. 
    A dominant peak indicates East-West oscillations with a period around 7 seconds. After the $16$\textsuperscript{th} of March 2022, the amplitude of the peak increased and its location (vertical lines) slightly shifted from $\zeta_{\max, \text{before}} \approx 1/7.8$ s$^{-1}$ to $\zeta_{\max, \text{after}} \approx 1/6.94$ s$^{-1}$.
    A similar analysis for the reference periods in 2021 shows almost no changes before and after the $16$\textsuperscript{th} of March.
    }
    \label{fig:fig3_pca_results}
\end{figure*}

Several periodic patterns occur in the grid frequency dynamics, most prominently the inter-area oscillations~\cite{Klein1991,Fritzsch2021}.
These modes are determined by the grid topology and may thus change strongly when a new grid is connected to an existing synchronous area.
This effect was intensively studied for the synchronization of the CE grid and Turkey~\cite{grebe2010low}.

We analyze potential changes in the spatio-temporal dynamics using synchronized frequency measurements at five different locations throughout the synchronous grid of Continental Europe provided by the Gridradar initiative \cite{gridradar}. 
The data set has been recorded with a temporal resolution of $100$ms and covers the time between the 16\textsuperscript{th} February and the 16\textsuperscript{th} of April 2022, i.e. one month before and after the synchronization. 
The specific measurement locations (i.e. Lleida in Spain, Herzogenrath and Bremen in Germany, Reisach in Austria and Sibiu in Romania) were selected to cover both the center and periphery of the synchronous grid of Continental Europe. 
We also stress the dependency on available data, which limits the free choice of measurements locations. 
As a reference data set, we also consider the same time period in the year 2021. 

To separate spatial and temporal patterns in the grid frequency signal, we employ Principal Component Analysis (PCA) \cite{bishop2006pattern}. The grid frequency $f(x,t)$ at location $x$ and time $t$ is decomposed as
\begin{align}
    f(x,t) = \sum_{n} A_n(t) v_n(x), 
\end{align}
where $n$ orders the different spatial components according to how much variability of the original signal they explain and $A_n(t)$ is the amplitude of component $n$ at time $t$. 
PCA chooses the components $v_n$ such that a maximum of the signal variability is explained by a minimum of components. 
Technically, the $v_n$ correspond to the eigenvectors of the covariance matrix and the associated eigenvalues $\tilde{\lambda}_1^f$ indicate the amount of variability explained by the respective component \cite{bishop2006pattern}.

Now, considering the locations $x \in$ [Lleida, Herzogenrath, Bremen, Reisach, Sibiu], the two leading principal components (i.e., $n=1,2$) are shown in Fig.~\ref{fig:fig3_pca_results}a-d. 
The first component $v_1(x)$ has very similar values at all locations $x$ and thus describes the bulk dynamics of the grid frequency. 
It is found that this bulk mode explains already above 99\% of the observed variance indicated by the eigenvalue $\tilde{\lambda}_1^f$, consistent with earlier results~\cite{rydingorjao2020open}. 
The second component $v_2(x)$ features opposite signs in the Western and Eastern part of the grid, i.e.~the frequency in the two parts always move in opposition. 
Thus, we can identify this component with the East-West inter-area mode of the grid.

We expect that the East-West mode would be most affected by the synchronization of the Ukrainian and Moldovan power grid, since this effective grid expansion strongly changed the grid structure at the Eastern end. To quantify potential changes, we compare the amplitude $A_2(t)$ before and after the $16$\textsuperscript{th} of March. The power spectral density  (Fig.~\ref{fig:fig3_pca_results}e) has a pronounced peak, corresponding to an oscillatory motion with a period around 7 seconds. We find that the amplitude of this peak increased after the $16$\textsuperscript{th} of March. Furthermore, the location of the peak slightly shifted from $\zeta_{\max, \text{before}} \approx 1/7.8$ s$^{-1}$ to $\zeta_{\max, \text{after}} \approx 1/6.94$ s$^{-1}$. Notably, the reference data set shows only very minor changes between the two respective time periods in 2021.

In conclusion, the data suggests that the East-West mode became slightly faster and more pronounced after the synchronization of the Ukrainian and Moldovan grid.

\section*{Cross border flows}

In addition to the changes in dynamics of the frequency, it is also expected that the power flows were affected by these recent events.
For one, power exchanges between Ukraine and Russia are known to have ended due to the start of the conflict.
In this context, we investigate the publicly available data of the power exchange between countries as recorded by the ENTSO-E and available on their transparency platform~\cite{entsoe}.
We focus on Ukraine and leave Moldova for future analysis. 

The data on the ENTSO-E platform distinguishes between countries which are mostly identical to the smallest unit in load-frequency control, i.e. control areas, with some exceptions:  Germany and Ukraine.
ENTSO-E only reports the power flow for existent connections between countries. 
Hence, we used this data set that describes the exchange between countries. 
Ukraine is a peculiar case, since it has three different regions that act distinctly different. 
The main UA-IPS grid that was emergency synchronized  to the CE-grid, UA-DobbTPP that is comprised of the larger Dobrotvir thermal power plant already has connections to the Polish grid, and UA-BEI comprised of the Burshtyn Energy Island located in the West of Ukraine. 

\begin{figure}
    \begin{center}
    \includegraphics[width=0.9\columnwidth]{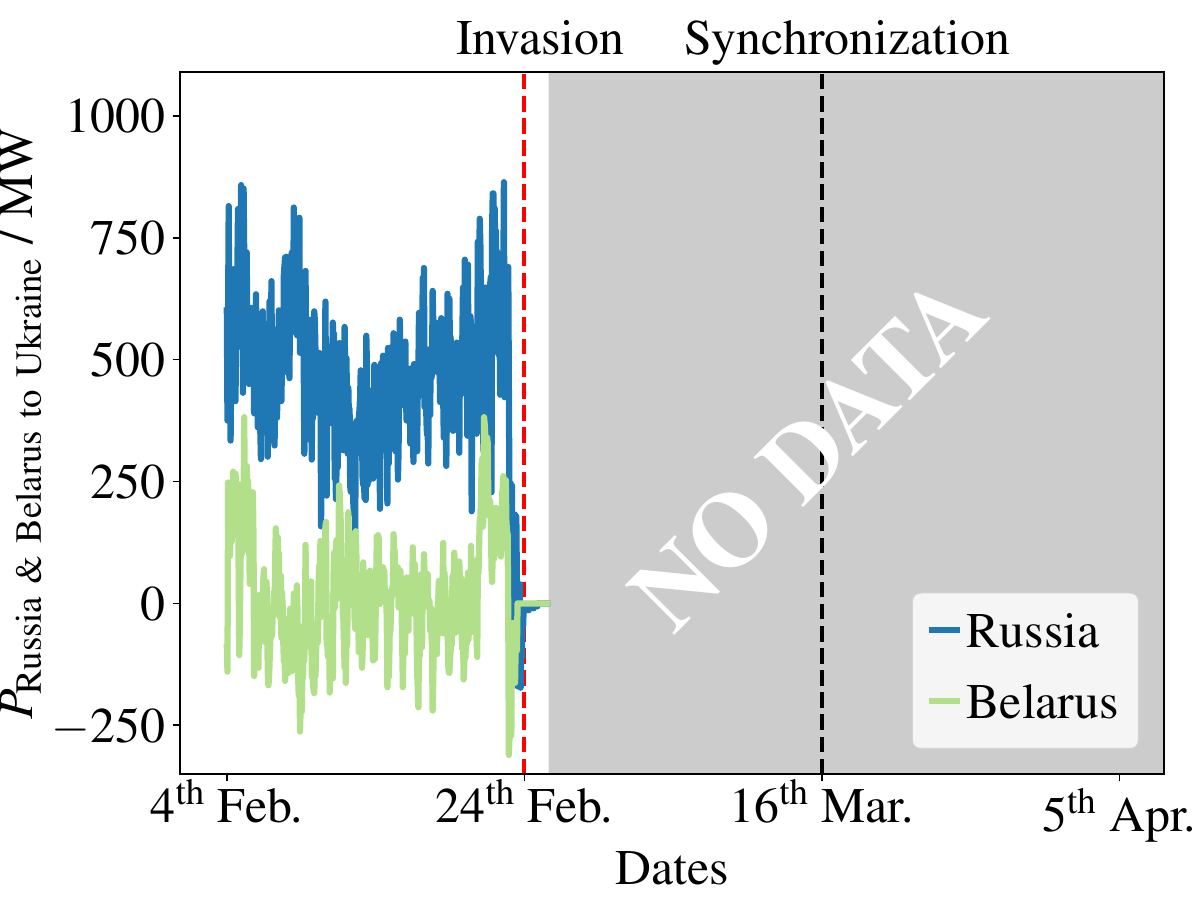}        
    \end{center}
    \caption{The war upended the flow of energy from Russia and Belarus to Ukraine. 
    We show the power flow from Russia and Belarus to Ukraine 20 days prior, between the invasion and the synchronization, and 20 days after, i.e., from the 4\textsuperscript{th} of February to the 5\textsuperscript{th} of April, 2022.
    As seen, no data are available in the ENTSO-E transparency platform after the date of the invasion~\cite{entsoe}.
    }
    \label{fig:4}
\end{figure}

Starting with the most obvious results, we analyze the exchange between Ukraine and both Belarus and Russia: Before the war, Ukraine imported power from Russia and Belarus but briefly after the war started, the flow came to a halt and no power was exchanged anymore, as seen in Fig.~\ref{fig:4}.
After 15:00 on February 25\textsuperscript{th}, no more data of power exchanges was reported.

\begin{figure}[tb]
    \begin{center}
    \includegraphics[width=0.9\columnwidth]{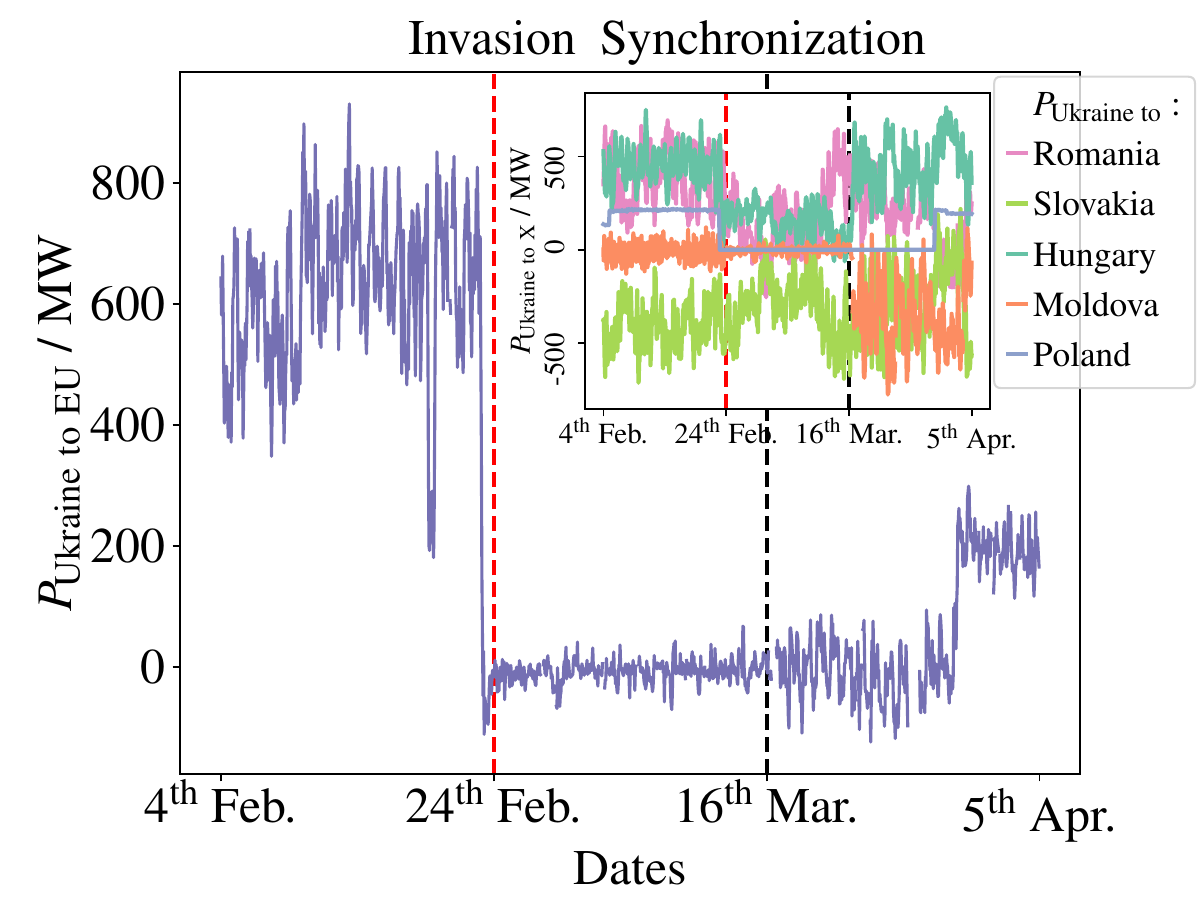}
    \end{center}
    \caption{The war lead to large changes in power exchange between Ukraine and Europe. 
    We show the power flow from Ukraine to Europe from the 4\textsuperscript{th} of February to the 5\textsuperscript{th} of April, 2022.
    This as abruptly ended at the outset of the conflict.
    In the inset we can see that some power was still being exported to Hungary and Romania, and some imported from Slovakia.
    The substantial difference is seen after the synchronization with the Continental European Grid, wherein large in-flows of power are passing from Moldova to Ukraine.}
    \label{fig:5}
\end{figure}

After Russia and Belarus stopped providing power to Ukraine, how did this impact power exchange to the rest of Europe?
As we see in Fig.~\ref{fig:5}, the power exchange with Ukraine's neighboring countries (Romania, Moldova, Slovakia, Hungary and Poland) reduced significantly.
In total the netflow even reversed to a slightly negative value and thus Ukraine at this point imported power, although very small amounts.
It is only after the synchronization with the Continental European grid that power is being exported into Europe once more.
The effect can clearly be seen once both Moldova and Ukraine join the Continental European Grid and we observe a clear inflow of power from Moldova to Ukraine, which was likely made possible by their synchronization with the Continental Europe power grid. 

\section*{Conclusion}

Within this article, we have presented an initial statistical analysis of how the emergency synchronization of the Ukrainian and Moldovan grids to the Continental European power grid has affected frequency dynamics and cross-border flows of electric power. 
We found evidence of increased fluctuations in the frequency, whereas the added inertia from Ukraine and Moldova should theoretically decrease the fluctuations.
Within the context of an ongoing war, it is noteworthy that while we observe an increase in fluctuations, this increase is small. 
One could have expected a much graver picture.
Further research will have to determine the exact source of these fluctuations.
While additional fluctuations might lead to additional control effort and thereby control costs, these additional operational costs are likely negligible compared to the influence of direct economic sanctions and high gas prices connected with the ongoing war~\cite{liadze2022economic, braesemann2022data}.
Finally, we noted the desired effect of the synchronization in the power flows: While the invasion stopped almost all imports and exports, the synchronization allowed flows between Europe and Ukraine.

Above all, let us emphasize the achievement performed by the ENTSO-E as well as the Ukrainian and Moldovan TSOs in synchronizing these three grids at such short notice and in such short time without causing any major disturbances.
It remains an open question how the synchronization of the Moldovan-Ukrainian power grid to the Continental European one will affect the long-term properties of the now synchronized area and whether the Russian invasion of Ukraine will lead to an accelerated synchronization of the Baltic power grids to the Continental European one.
Another open question is how the very recent (October 2022) Russian attacks on the Ukrainian power system,  affect the frequency statistics of Figs.~\ref{fig:1}--\ref{fig:fig3_pca_results}.

\section*{Data availability} All source code and data for one location (Bremen) can be found online~\cite{Boettcher2022}. 
Further data are not disclosed openly by Gridradar due to the potential misuse by third parties, e.g. the intent to damage critical infrastructure. Access to the data for scientific purposes is possible by contacting Gridradar directly.
For information on the data collection, associated measurements units, and technical aspects of power-grid frequency control, see \href{https://gridradar.net/en}{https://gridradar.net/}~\cite{gridradar}. For details on power flows between countries in the ENTSO-E area, see \href{https://transparency.entsoe.eu/}{https://transparency.entsoe.eu/}~\cite{entsoe}.

\section*{Acknowledgements} 
We thank Tobias Veith from Gridradar for providing us with frequency data.
We gratefully acknowledge support from the German Federal Ministry of Education and Research with grant no.~03EK3055B, the Helmholtz Association via the grant \textit{Uncertainty Quantification -- From Data to Reliable Knowledge (UQ)} and grant no.~VH-NG-1727, and the Deutsche Forschungsgemeinschaft (DFG,   German Research Foundation) with grant no.~ 491111487.

\bibliography{bib}

\end{document}